\documentclass[anon,12pt]{l4dc2021}

\DontPrintSemicolon

\usepackage{soul}
\usepackage{graphicx}      
\usepackage{booktabs}
\newcommand{\ra}[1]{\renewcommand{\arraystretch}{#1}}
\usepackage{glossaries-prefix}
\newacronym{pid}{PID}{proportional-integral-derivative}
\newacronym[prefixfirst={a\ },
            prefix={an\ }
            ]{ml}{ML}{machine learning}
\newacronym{drl}{DRL}{deep reinforcement learning}
\newacronym[prefixfirst={a\ },
            prefix={an\ }
            ]{nn}{NN}{neural network}
\newacronym[prefixfirst={a\ },
            prefix={an\ }
            ]{rl}{RL}{reinforcement learning}
\newacronym[prefixfirst={a\ },
            prefix={an\ }
            ]{mdp}{MDP}{Markov decision process}
\newacronym[prefixfirst={a\ },
            prefix={an\ }
            ]{mpc}{MPC}{model predictive control}
\newacronym{nmpc}{NMPC}{nonlinear model predictive control}
\newacronym{empc}{EMPC}{economic model predictive control}
\newacronym[prefixfirst={a\ },
            prefix={an\ }
            ]{mpcr}{MPC}{model predictive controller}
\newacronym[prefixfirst={a\ },
            prefix={an\ }
            ]{lqr}{LQR}{linear quadratic regulator}
\newacronym[prefixfirst={a\ },
            prefix={an\ }
            ]{lqg}{LQG}{linear qudratic gaussian}
\newacronym{fpga}{FPGA}{field-programmable gate array}
\newacronym{soc}{SoC}{state of charge}
\newacronym{ocp}{OCP}{optimal control problem}
\newacronym{ou}{OU}{Ornstein–Uhlenbeck}
\newacronym{uav}{UAV}{unmanned aerial vehicle}

\usepackage{textcomp}
\usepackage{bbold}

\definecolor{seb}{rgb}{0.8,1,0.8}
\newcommand{\seb}[1]{ \begin{center}
\fcolorbox{seb}{seb}{\parbox[t]{0.9\linewidth}{\textbf{Seb:} #1}} \end{center}}

\newcommand{\pluseq}{\mathrel{{+}{=}}}

\newcommand {\matr}[2]{\left[\begin{array}{#1}#2\end{array}\right]}

\title[Optimization of the MPC Update Interval Using RL]{Optimization of the Model Predictive Control Update Interval Using Reinforcement Learning}
\usepackage{times}

\author{%
 \Name{Eivind Bøhn} \Email{eivind.bohn@sintef.no}\\
 \addr Department of Mathematics and Cybernetics, SINTEF Digital, Oslo, Norway
 \AND
 \Name{Sebastien Gros} \Email{sebastien.gros@ntnu.no}\\
 \addr Department of Engineering Cybernetics, NTNU, Trondheim, Norway
 \AND
 \Name{Signe Moe} \Email{signe.moe@sintef.no}\\
 \addr Department of Mathematics and Cybernetics, SINTEF Digital, Oslo, Norway
 \AND
 \Name{Tor Arne Johansen} \Email{tor.arne.johansen@ntnu.no}\\
 \addr Centre for Autonomous Marine Operations and Systems, Department of Engineering Cybernetics, NTNU, Trondheim, Norway%
}

\begin{document}

\maketitle

\begin{abstract}
In control applications there is often a compromise that needs to be made with regards to the complexity and performance of the controller and the computational resources that are available. For instance, the typical hardware platform in embedded control applications is a microcontroller with limited memory and processing power, and for battery powered applications the control system can account for a significant portion of the energy consumption. We propose a controller architecture in which the computational cost is explicitly optimized along with the control objective. This is achieved by a three-part architecture where a high-level, computationally expensive controller generates plans, which a computationally simpler controller executes by compensating for prediction errors, while a recomputation policy decides when the plan should be recomputed. In this paper, we employ \gls*{mpc} as the high-level plan-generating controller, a linear state feedback controller as the simpler compensating controller, and \gls*{rl} to learn the recomputation policy. Simulation results for two examples showcase the architecture's ability to improve upon the \gls*{mpc} approach and find reasonable compromises weighing the performance on the control objective and the computational resources expended.
\end{abstract}

\glsresetall

\section{Introduction}\label{sec:introduction}
\Gls*{rl} is a class of machine learning algorithms that discover optimal policies in the sense of maximizing some notion of utility through interacting with the dynamic system to be controlled \citep{sutton_reinforcement_2018}. It therefore offers an interesting complementary approach to developing control systems for problems unsuited to traditional control approaches based purely on first principles modeling. \Gls*{rl} has already proven to be the state-of-the-art approach to classes of problems requiring complex decision-making over large time frames such as game playing \citep{mnih2015human,Schrittwieser2019MasteringAG,berner2019dota}. However there are some major limitations to the \gls*{rl} framework that inhibits its applicability to control applications, notably the lack of guarantees for safe operation and ability to specify constraints. Some of these limitations can be addressed by combining \gls*{rl} with trusted control approaches such as \gls*{mpc}, thereby allow for learning of various aspects of the control problem while having a trusted controller oversee the process and ensure safe operation \citep{zanon_safe_2020,gros_safe_2020}.

On the other hand, the \gls*{mpc} approach has a few major drawbacks of its own, notably its reliance on having a reasonably accurate model of the system. Unfortunately, obtaining such a model can be both difficult and costly. Moreover, the \gls*{mpc} approach has a heavy online computational cost, as it involves solving an \gls*{ocp} at every time step over a selected horizon. \gls*{mpc} is therefore in many cases unsuited for low-powered or battery-driven applications, and employing \gls*{mpc} in these scenarios often necessitates some form of compromise, see e.g. \cite{le_approximate_2012} and \cite{gondhalekar_tackling_2015}. While we in this paper address this problem from an energy perspective of reducing the frequency of computations, one can also look at reducing the complexity of the \gls*{mpc} computation itself. The two main tunable parameters in this regard is the optimization horizon length, which when lowered can lead to myopic behaviour, and the step size which when too large impacts control performance as the controller is unable to properly respond to high frequency dynamics and transient disturbances. The main techniques to reduce computational complexity are early termination (suboptimal) \gls*{mpc} \citep{scokaert_suboptimal_1999}, explicit \gls*{mpc} \citep{bemporad_explicit_2002}, semi-explicit \gls*{mpc} \citep{goebel_simple_2015} and move-blocking \citep{cagienard_move_2004}.

As an alternative to the normal digital control system approach of measuring system state and calculating control signal at equidistant points in time, one can instead select these points by some other criteria, yielding a control system that consists of a control law and a triggering policy. This event-triggered control paradigm could be used to increased control performance or reduce frequency of computations based on how the triggering policy is designed \citep{astrom_comparison_2002,heemels_analysis_2008}. Event-triggered \gls*{mpc} has been suggested in the literature to reduce resource usage particularly in networked communication systems and multi-agent systems \citep{iino_event_2009,berglind_self_2012,li_event_2014,chakrabarty_event_2018}. In these works the triggering policy is a predetermined hand-crafted policy based on domain knowledge about the system and the controller, a learned policy on the other hand can be derived without requiring system knowledge and be adaptive to changing dynamics and uncertain disturbances. In \cite{yoo_event_2019} the authors use a learned empirical risk minimization model to predict the unknown system noise in \pgls*{mpc} framework and use its output to determine the triggering thresholds. For a recent review of machine learning applied to event-triggered networked control systems, see \cite{sedghi2020machine}. 

In this work we look at the single agent setting and suggest to learn the triggering policy with \gls*{rl}. Several other works such as \cite{baumann_deep_2018} and \cite{yang_event_2019} have proposed \gls*{rl} for learning the triggering policy, however these works simultaneously learn the control law as opposed to our work which employs \gls*{mpc} for this purpose. The contribution of this paper lies in proposing a novel architecture involving a dual mode \gls*{mpc} and \gls*{lqr} control law and \pgls*{rl} based triggering policy, and in deriving how the event-triggered \gls*{mpc} problem can be framed as \pgls*{mdp}. The \gls*{mpc} solves the \gls*{ocp}, providing a plan in the form of an input sequence and the predicted state trajectory from executing the input sequence, while the \gls*{lqr} provides an additive compensatory input based on the state trajectory prediction errors extending the viability of \gls*{mpc} solution. Finally, the \gls*{rl} triggering policy, henceforth referred to as the recomputation policy, selects time instants when the improved control performance from recomputing the \gls*{mpc} solution outweigh the computational costs. We empirically demonstrate the effectiveness of the proposed architecture through two simulated use cases.

The rest of the paper is organized as follows. First, Section \ref{sec:theory} presents the requisite theoretical background of the proposed architecture, while Section \ref{sec:method} presents the implementation details. Section \ref{sec:experiments} describes the two case studies illustrating the proposed system, and Section \ref{sec:results} presents the results of these case studies and discusses their significance. Finally, Section \ref{sec:conclusion} presents our closing thoughts on the paper.

\section{Preliminaries and Problem Formulation}\label{sec:theory}
\subsection{Model Predictive Control}
\gls*{mpc} is a control technique where the current control action is obtained by solving at each time step an open loop finite horizon \gls*{ocp} \eqref{eq:mpc}, where the initial state is given as the current state of the plant. The \gls*{mpc} solution consists of an input sequence minimizing the objective function over the optimization horizon, of which only the first input is applied to the plant, and the \gls*{ocp} is then computed again at the subsequent time step. We consider in this paper discrete time direct state feedback \gls*{mpc}, meaning the \gls*{mpc} gets exact measurements of the state of the plant at discrete time intervals, and the optimization variables are assumed constant between sampling instances.

\begin{subequations}\label{eq:mpc}
    \begin{align}
        \min_{x, u} \quad &\sum_{k=0}^{N-1}\left[\ell(x_k, u_k, \hat{p}_k) + \Delta u_k^\top D \Delta u_k \right] + m(x_N), \label{eq:mpc:obj}  \\ 
        \textrm{s.t. \quad} &x_0 = \bar{x} \label{eq:mpc:init} \\ 
        &x_{k+1} = \hat{f}(x_k, u_k, \hat{p}_k) \label{eq:mpc:dynamics} \\ 
        &H(x_k, u_k) \leq 0 \label{eq:mpc:H}
    \end{align}
\end{subequations}
Here, $x_k$ is the plant state vector at time $k$ and $\bar{x}$ is the measurement of these states, $u_k$ is the control input vector and $\Delta u_k = u_k - u_{k-1}$, $\hat{p}_k$ are time-varying parameters whose value is forecasted over the optimization horizon, $\hat{f}$ is the \gls*{mpc} model dynamics which may differ from the plant system dynamics $f$ due to unmodeled effects such as process noise $w_k$, $H$ is the constraint vector and $N$ is the optimization horizon length. The objective function consists of the stage cost $\ell(x_k, u_k, \hat{p}_k)$, which --- in addition to the input-change term discouraging bang-bang control --- is summed over every step of the optimization horizon except the last, and the end of horizon cost $m(x_N)$. The state and control inputs are subject to constraints, which must hold over the whole optimization horizon for the \gls*{mpc} solution to be considered feasible.

We propose in this paper to modify the traditional \gls*{mpc} approach by not statically recomputing the \gls*{mpc} solution at every time step, but instead to utilize a recomputation policy to select at each time step whether or not to recompute the \gls*{mpc} solution. This means that of the \gls*{mpc} input sequence $u_{k,\dots,k+N-1}$ computed at time $k$, not only the first input $u_k$  is applied but rather a variable number of inputs $u_{k, \dots, k + n}, \enspace n < N$  are applied sequentially to the plant at the corresponding time instance $i, \enspace k \leq i \leq k + n$, until the recomputation policy triggers the recomputation of the \gls*{mpc} solution at time $k + n$. We detail this recomputation policy next.

\subsection{Event-triggered MPC}
The recomputation policy $\pi_\theta(a \thinspace | \thinspace s)$ is a stochastic policy parameterized by $\theta$, defining a probability distribution over a binary output action, $a$, where $a_k = 1$ corresponds to recomputing the \gls*{mpc} solution at step $k$ and $a_k = 0$ corresponds to not recomputing it. The recomputation policy $\pi_\theta$ receives a state $s$ that is not simply the real system state $x$ because $\pi_\theta$ depends on the state of the system when the \gls*{mpc} was last recomputed. For classical \gls*{rl} theory to apply for $\pi_\theta$ the state $s$ must be a Markov state. We will therefore define next an augmented state space $\mathcal{S}$ that does have this property.

Assume that the \gls*{mpc} solution delivers an input sequence $u_{k,\ldots,k+N-1}$ associated to a measured state $\bar{x}_k$, at time instant $k$ and that the sequence $u_{k,\ldots,k+n}$ is applied to the system. The inputs received by the plant in the time interval $k,\ldots,k+n$ are then a function of state $\bar x_k$ at time $k$. Since $n\in \left\{k,\ldots,N-1\right\}$ is not known a priori, one ought to generally view the control system stemming from the triggering policy and the \gls*{mpc} scheme as being a control law from an augmented state space, $\mathcal{S}$, containing 1. the current state of the system, 2. the state of the system when the last optimization took place, and 3. the number of time samples from which the last optimization took place. Labelling the current time of the system as $i$ and the last time when the optimization occurred as $k$, that augmented state reads as:
\begin{align}
s = \matr{c}{\tilde x_{i} \\ \tilde x_k  \\ i-k}
\end{align}
where $k \leq i \leq k+N-1$, and where $\tilde x_{i}$ has the real system dynamics, and where the deterministic state transition $\tilde x_k \leftarrow \tilde x_{i},\enspace k \leftarrow i$ occurs when $a_i = 1$ is drawn from the stochastic policy $\pi_\theta(a_i \thinspace | \thinspace s_i)$. The \gls*{mpc} control law actually deployed on the system then reads as:
\begin{align}
    \pi^\mathrm{MPC}(s) = u^{\mathrm{MPC}}_{i-k}(\tilde x_k)
\end{align}
where $u^{\mathrm{MPC}}_{i-k}$ is the $i-k^\mathrm{th}$ element of the \gls*{mpc} solution solved for the initial state $\tilde x_k$. 

The recomputation policy, and the \gls*{mpc} and \gls*{lqr} control laws together form a control system defining the inputs actually applied to the plant in a closed loop system. The control system and the plant dynamics together define the state transition dynamics in the augmented state space $\mathcal{S}$, where states $s$ does have the Markov property. The event-triggered \gls*{mpc} problem is therefore an \gls*{mdp} in the augmented state space $\mathcal{S}$, such that classic \gls*{rl} can be applied on $s$. In that context external parameters (forecasts) can be seen as being part of the states $\tilde x_k$, $\tilde x_i$.

\subsection{Reinforcement Learning}
The closed loop system \gls*{mdp} as described above is defined by the tuple $\langle \mathcal{S}$, $\mathcal{A}$, R, $\mathcal{T}$, $\mathcal{T}_0$, $\gamma \rangle$.  $\mathcal{S}$ is the set of states $s$, and $\mathcal{A} = \{0, 1\}$ is the set of actions. $\mathcal{T} : \mathcal{S} \times \mathcal{A} \rightarrow \mathcal{S}$ is the discrete-time transition function describing the probabilistic transitions between states as a function of time and actions, and $\mathcal{T}_0$ is a distribution of initial states. $R : \mathcal{S} \times \mathcal{A} \rightarrow \mathbb{R}$ is the cost function which assigns a scalar value to the states of the problem. Lastly, $\gamma \in \left[0, 1\right)$ is the discount factor weighing the relative importance of immediate and future costs. 



We consider the episodic setting, where an episode consists of a sequence of states and actions of length $T$, denoted by $\tau = (s_0, a_0, s_1, a_1, \dots, s_T)$. We further define the return $G(\tau) = \gamma^T R(s_T) + \sum_{k=0}^{T-1} \gamma^k R_k(s_k, a_k)$ 
as the total cost of the episode, where $R(s_T)$ is a terminal state cost. \gls*{rl} encompasses a set of algorithms to generate policies $\pi : \mathcal{S} \rightarrow \mathcal{A}$ with the goal of finding the optimal policy $\pi^*$ that optimizes the \gls*{rl} objective which in this setting can be formulated as \eqref{eq:rl:J}. 

\begin{subequations}
    \begin{align}
        J(\pi) = \mathbb{E} \left[G(\tau) \thinspace | \thinspace \pi\right] \label{eq:rl:J} \\
        \pi^* = \max_\pi J(\pi)
    \end{align}
\end{subequations}

The expectation in \eqref{eq:rl:J} is taken over the initial state distribution, the probabilistic transition function, and the possibly stochastic policy $\pi$. 

\subsubsection{Policy Gradients}
Policy gradient algorithms is a family of \gls*{rl} methods that directly optimize the parameters $\theta$ of parameterized policies $\pi_\theta$ by estimating the gradient of the policy performance index $J$ with respect to $\theta$, and using a gradient descent scheme to minimize the objective \eqref{eq:rl:J}. The policy gradient is given by \citep{sutton_reinforcement_2018}:
\begin{subequations}
    \begin{align}
        \nabla_\theta J(\pi_\theta) &= \mathbb{E}\left[A_{\pi_\theta}(\tau) \nabla_\theta \log\pi_\theta(\tau) \right] \\
        \theta &\leftarrow \theta + \alpha \nabla_\theta J(\pi_\theta)
    \end{align}
\end{subequations}
Where $A_{\pi_\theta}(\tau)$ is the advantage function describing the relative value of states with respect to the costs obtained over the future trajectory generated by $\pi_\theta$ from the state in question. $A(\tau)$ is often replaced by other functions of the costs depending on the algorithm, e.g. by $G(\tau)$ for ``vanilla" policy gradient. Policy gradient algorithms can therefore intuitively be interpreted as adjusting the log likelihood of actions such that actions that lead to lower costs are more likely to be chosen. This gradient relies on access to the costs for the states and actions in the trajectory, which is obtained by sampling trajectories from the plant. Policy gradient methods have good convergence guarantees in theory, but since it is a sampling based approach it suffers from high variance in the gradient estimates in practice \citep{peters_reinforcement_2008}.

\subsubsection{Policy Representation}
As the action space of the policy $\pi_\theta$ in our method is binary we represent the policy as a logistic regression model. In the logistic model, the logarithm of the odds of a binary dependent variable taking on a given value is modeled as a linear combination of the input variables \eqref{eq:logreg_probs}:

\begin{align}
        &\pi_\theta(0 \thinspace | \thinspace s) = \frac{1}{1 + e^{-\theta^\intercal s}}, \enspace \pi_\theta(1 \thinspace | \thinspace s) = 1 - \pi_\theta(0 \thinspace | \thinspace s) \label{eq:logreg_probs}
\end{align}

\section{Proposed Control System Architecture}\label{sec:method}

The proposed architecture is described in Algorithm \ref{alg:cfs}. The \gls*{mpc} solution is computed at the first time step, generating an optimal predicted state trajectory $\hat{x}_{k+1, \dots, k + N}$ and a sequence of optimal inputs that produces this trajectory $u^{\mathrm{MPC}}_{k, \dots, k + N - 1}$. The first input is then applied to the system as usual, but instead of discarding the rest of the \gls*{mpc} solution, we allow a learned policy to decide based on an observation of the current recomputation state, $s$, whether the last \gls*{mpc} solution is still of acceptable quality, or if it should be recomputed. If the recomputation policy decides not to recompute, a tracking \gls*{lqr} is applied to the state prediction error $\epsilon_x$ to correct the state trajectory to match the plan generated by the \gls*{mpc}. Algorithm \ref{alg:cfs} describes the operation of the control system with a fixed recomputation policy, e.g. after the \gls*{rl} policy is done learning, as how and when the learning is performed would depend on the chosen learning algorithm.

In this paper we use the GPOMDP algorithm \citep{Baxter_2001} to train the \gls*{rl} recomputation policy, and implement the \gls*{mpc} with the CasADi framework \citep{Andersson2019}, the Ipopt optimizer \citep{ipopt} and the do-mpc python library \citep{lucia2017rapid}.
\begin{remark}
In our experiments computing the \gls*{lqr} input and evaluating the \gls*{rl} policy takes on the order of $10^{-3}$ less time than computing the \gls*{mpc} solution, lending credence to this architecture saving significant computational resources.
\end{remark}

\begin{algorithm2e}[htbp]
    \SetAlgoLined
    \caption{Control System Architecture}

    Initialize state and parameters: $x_k = x_0, \enspace \hat{p}_k = p_0$ \;
    \For{$k = 0, 1, 2, \dots$}{
        Compute MPC solution: $\hat{x}_{k+1, \dots, k+N}, \enspace u^{\mathrm{MPC}}_{k, \dots, k + N - 1} = \mathrm{MPC}(x_k, \hat{p}_k)$ \;
        Execute MPC control input: $u^{\mathrm{MPC}}_{k}$  \;
        \For{i = k, \dots, k + N - 1}{
            Wait for next sampling instant and measure system state: $\bar{x}_{i}$ \;
            Compute prediction errors: $\epsilon_{x_{i}} = \hat{x}_{i} - \bar{x}_{i}$ \;
            \eIf{\textup{Recomputation policy} $\pi(a_i \thinspace | \thinspace s_i)$ \textup{ draws } $a_i = 1$}{
                Break loop
            }{
                Compute additive LQR input: $u_{i} = u^{\mathrm{MPC}}_{i} + LQR(\epsilon_{x_{i}})$ \;
                Apply input constraints: $u_{i} = \min\left(\max\left(u_{i}, u^{\mathrm{low}}\right), u^{\mathrm{high}}\right)$ \;
                Execute control input: $u_{i}$
            }
        }
        Update MPC model, state and parameters: $k \leftarrow i$
    }
    \label{alg:cfs}
\end{algorithm2e}

\section{Case Studies}\label{sec:experiments}
\subsection{Systems}

\subsubsection{Cart Pendulum}
We first look at a nonlinear system where the quality of the \gls*{mpc} control input is considerably better than that of the \gls*{lqr}. For this purpose we use the well studied classical control task of balancing an inverted pendulum mounted on a small cart. This system represents the problem of embedded control of battery powered systems where the \gls*{mpc} is needed for sufficient control performance but its computation accounts for a substantial part of the total battery resources available.

The state space consists of $x_k = \left[\eta, v, \beta, \omega \right]^\intercal$ which is position and velocity of cart along horizontal axis, and angle of pendulum to the vertical axis and angular velocity of the pendulum, respectively. We discretize the system equations in (\ref{eq:cartpend:pos}-\ref{eq:cartpend:lqr}) with step time $k = 0.1s$, and add unmeasured process disturbance $w_k = \left[0, 0, 0, \mathcal{N}(0, 1)\right]^\intercal$, which means that feedback control from the \gls*{lqr} is a necessary addition to the \gls*{mpc} plan to stabilize the pendulum. We sample initial states according to $x_0 = \left[0, \mathcal{U}(-1, 1), \mathcal{U}(-0.78, 0.78), \mathcal{U}(-1, 1)\right]^\intercal$, where $\mathcal{U}$ is the uniform distribution, and put constraints on the input $-25 \leq u_k \leq 25$, which is a force applied to the cart along the horizontal axis. Finally, the physical parameters are the pendulum length and weight, $l = 1, \enspace m=0.1$, the total weight of the cart and the pendulum, $M = 1.1$, and the gravitational acceleration, $g = 9.81$. 

The \gls*{mpc} is configured with $N = 20,\thinspace D = 10^{-4},\thinspace \ell(x_k, u_k, p_k) = \beta_k^2,\thinspace m(x_N) = \beta_N^2$. This objective function promotes stabilization of the pendulum, while the \gls*{rl} policy cost function $R_k(s_k, a_k) = \beta_k^2 + 5\cdot10^{-5} a_k$ has an additional objective to minimize computation. The \gls*{lqr} is based on a fixed linearized model \eqref{eq:cartpend:lqr} obtained by using the small angle approximation, while its cost matrices are $Q_{lqr} = diag(0, 1, 10, 10),\thinspace R_{lqr} = 0.1$.

\begin{subequations}
    \begin{align}
        \dot{\eta} &= v \label{eq:cartpend:pos}, \enspace \dot{v} = \frac{m g sin(\beta) cos(\beta) - \frac{4}{3}(u + m  l  \omega ^ 2 sin(\beta))}{m cos^2(\beta) - \frac{4}{3} M} \\
        \dot{\beta} &= \omega, \enspace \dot{\omega} = \frac{M  g  sin(\beta) - cos(\beta)  (u + m  l  \omega ^ 2  sin(\beta))}{\frac{4}{3} M  l - m  l  cos^2(\beta)} \label{eq:cartpend:omega} \\
        \dot{\epsilon}_x &= \begin{bmatrix}
        0 & 1 & 0 & 0 \\
        0 & 0 & -\frac{m g}{4/3 M - m} & 0 \\
        0 & 0 & 0 & 1 \\
        0 & 0 & \frac{M g}{l(4/3 M - m)} & 0
        \end{bmatrix} \epsilon_x + \begin{bmatrix}
        0 \\
        \frac{1}{M - 3/4 m} \\
        0 \\
        \frac{-1}{l (4/3 M - m)}
        \end{bmatrix} u^{\mathrm{LQR}} \label{eq:cartpend:lqr}
    \end{align}
\end{subequations}

\subsubsection{Battery Storage}

The battery storage system models the trading of electricity from a battery storage that is subject to quickly changing electricity prices as well as uncertain production and consumption from the electricity grid it is connected to \eqref{eq:battery}. Here, $x_k$ is the state of charge, i.e. the fraction of available battery capacity, $C = 10$ kWh is the total battery capacity, $u_k$ is the trading of power from the battery on the market where $u_k > 0$ represents selling and $u_k < 0$ is buying. $P_k$ is the sum of production and consumption on the power grid, and $\lambda_{e_k}$ is the time-varying price of electricity. 

This is an economic problem where the objective is to maximize the profit of the battery storage through the sum of trading $\ell(x_k, u_k, p_k) = \lambda_{e_k} u_k$ and the potential of the storage $m(x_N) = \bar{\lambda_e} C x_N, \thinspace R_k(s_k, a_k) = \lambda_{e_k} u_k \delta + \bar{\lambda_e} C x_T$, where $\bar{\lambda_e} = \mathbb{E}[\lambda_e]$ is the average price of electricity. The \gls*{mpc} has an optimization horizon of 20 time steps, $N = 20$. In this problem we look at fast trading, i.e. $\delta = 10s$, which is dominated by the balance market. Spot prices are evaluated every hour and fixed for the duration, while the prices are highly stochastic and vary around the spot prices on the balance market due to e.g. unforeseen errors on the grid. The control system receives forecast for the future production and consumption as well as the expected price of electricity. It then needs to plan whether to save up electricity for better times ahead or sell surplus charge for a good price. We assume that when the \gls*{mpc} solution is computed it runs for the whole sampling interval on a computer consuming $\eta_c = 100$W from the battery, modeled through the $\eta_c a_k$ term in \eqref{eq:battery}. Lastly, only 10\% of the battery capacity can be traded in a single window \eqref{eq:battery:c_u}, with no cost for changing amount between windows, $D = 0$. The \gls*{lqr} accounts for errors in the forecasted production and consumption, as well as the consumption from the \gls*{mpc} computation itself which is not part of the \gls*{mpc} model \eqref{eq:battery:lqr}, $Q_{lqr} = 1, \thinspace R_{lqr} = 0.1$.

\begin{figure}[htbp]
\floatconts
    {fig:forecasts}
    {\caption{Simulated values for production and electricity prices in the battery storage system, along with forecasts generated by \gls*{rl} policy and the every t = 20 baseline policy. The crosses mark an \gls*{mpc} computation, showing how forecast accuracy decreases with time.}}
    {\includegraphics[width=1\textwidth]{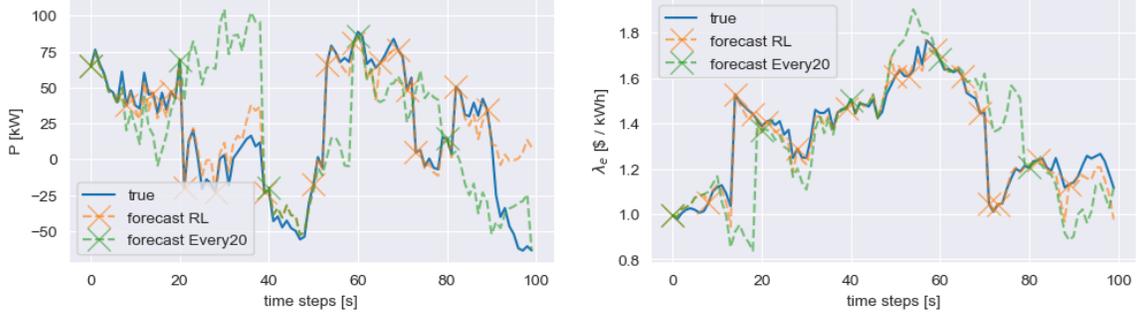}}
\end{figure}

The uncertainty in this problem lies in the forecasts of the time-varying parameters, $P_k$ and $\lambda_ {e_k}$. These grow more unreliable the further ahead in time we forecast, and as such the plan generated by the \gls*{mpc} becomes less and less optimal. The true values of the time-varying parameters are generated by sampling a base value with varying duration, representing the changing power demands due to events as described above, with an additive \gls*{ou} process. Further, forecasts are generated by assuming the base value is constant over the forecast length and adding a second \gls*{ou} noise process whose strength is linearly scaled with forecast length, see figure \ref{fig:forecasts}. 

\begin{remark}
$P$ and $\lambda_e$ are not based on real historical data, and are generated with the aim of emulating the market as described above and to produce a non-trivial test problem.
\end{remark}

\begin{subequations}
    \begin{align}
        x_{k+1} &= x_k - \frac{\delta}{C}(u_k - P_k - \eta_c a_k), \enspace 0 \leq x_k \leq 1  \label{eq:battery} \\
        -\frac{0.1C}{\delta} &\leq u_k \leq \frac{0.1C}{\delta} \label{eq:battery:c_u} \\
        \epsilon_{x_{k+1}} &= \epsilon_{x_k} -\frac{1}{C} \delta u^{\mathrm{LQR}}_k \label{eq:battery:lqr}
    \end{align}
\end{subequations}

\subsection{Baseline Policies}
To assess the quality of the \gls*{rl} policy we compare the learned policy to the fixed baseline policies: standard \gls*{mpc} ($\pi(a_k \thinspace | \thinspace s_k) \equiv 1)$, never recomputing until end of horizon ($\pi(a_k \thinspace | \thinspace s_k) \equiv 0$), and recomputing on a fixed static schedule every $t$ time steps ($\pi^t(a_k \thinspace | \thinspace s_k) = (k \mod t = 0))$.

\begin{remark}
We experimented with early termination \gls*{mpc} based on maximum allowed iterations as well as optimality tolerance, but its performance was considerably worse than the other approaches and as such it is not included. Early termination would likely perform better with a sequential quadratic programming (SQP) based optimizer that exploits warm start more efficiently than Ipopt.
\end{remark}

\subsection{Training and Evaluation}
We train and evaluate the policies in an episodic setting, where an episode consists of 100 time steps with randomly generated initial conditions, process-noise, time-varying parameters etc. as discussed above for each system. The \gls*{rl} state $s$ is augmented with a squared version of each component, and the whole state vector is normalized before being fed to the policy. To evaluate the policies and minimize the effects of the random variables on the results, we construct a test set for each system consisting of 100 episodes where all random variables are drawn in advance such that the episode is consistent across policy evaluations. To account for the inherent stochasticity of the \gls*{rl} policy, we evaluate the policy five times over the test set and report mean results. We use the undiscounted return $G(\tau), \enspace \gamma = 1$ as the objective value to compare policies on the episode $\tau$ and average these returns over the test set.

As the objective of the policy is to learn when it is viable to not recompute the \gls*{mpc} solution, we initialize the \gls*{rl} policy to (with high probability) mimic the standard \gls*{mpc} approach and always elect to recompute. For both systems we train with $\gamma = 0.975$ and a learning rate of $\alpha = 0.05$.

\section{Results and Discussion}\label{sec:results}

\begin{figure}[htbp]
\floatconts
    {fig:training}
    {\caption{Learning curves for the training phase of the \gls*{rl} recomputation policy showing mean values and standard deviation (shaded region) over five different seeds.}}
    {%
    \subfigure[Cart pendulum system]{%
        \label{fig:training:cartpend}
        \includegraphics[width=0.6975\textwidth]{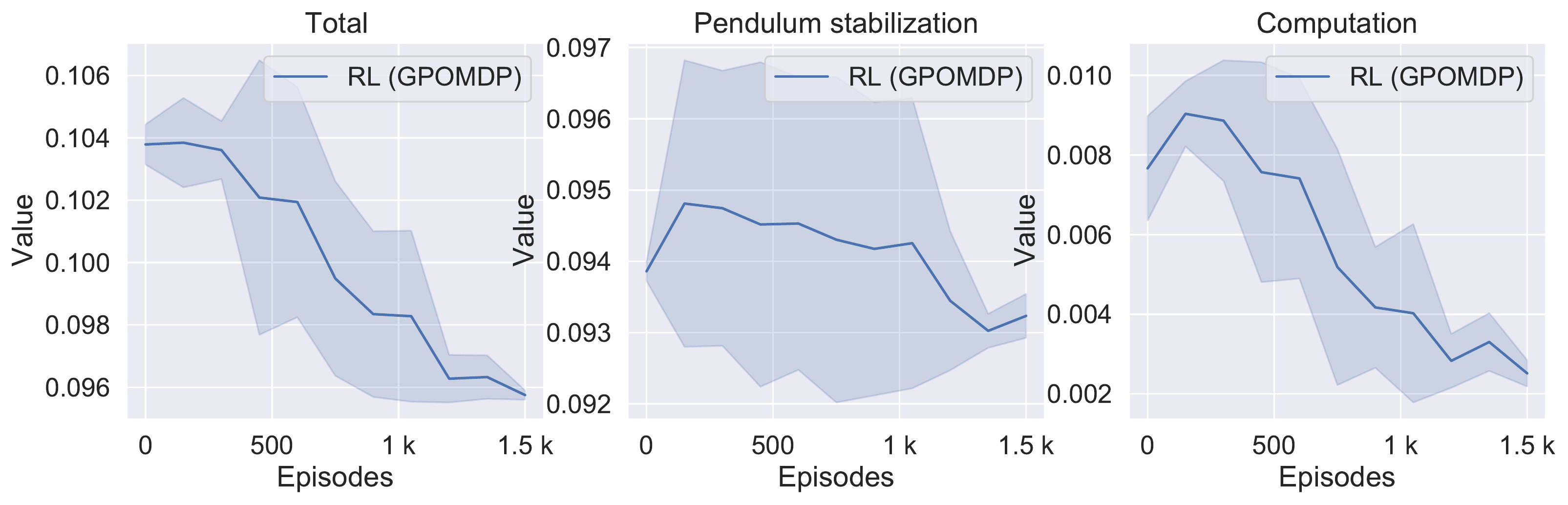}
    }
    \subfigure[Battery storage system]{%
        \label{fig:training:battery}
        \includegraphics[width=0.2825\textwidth]{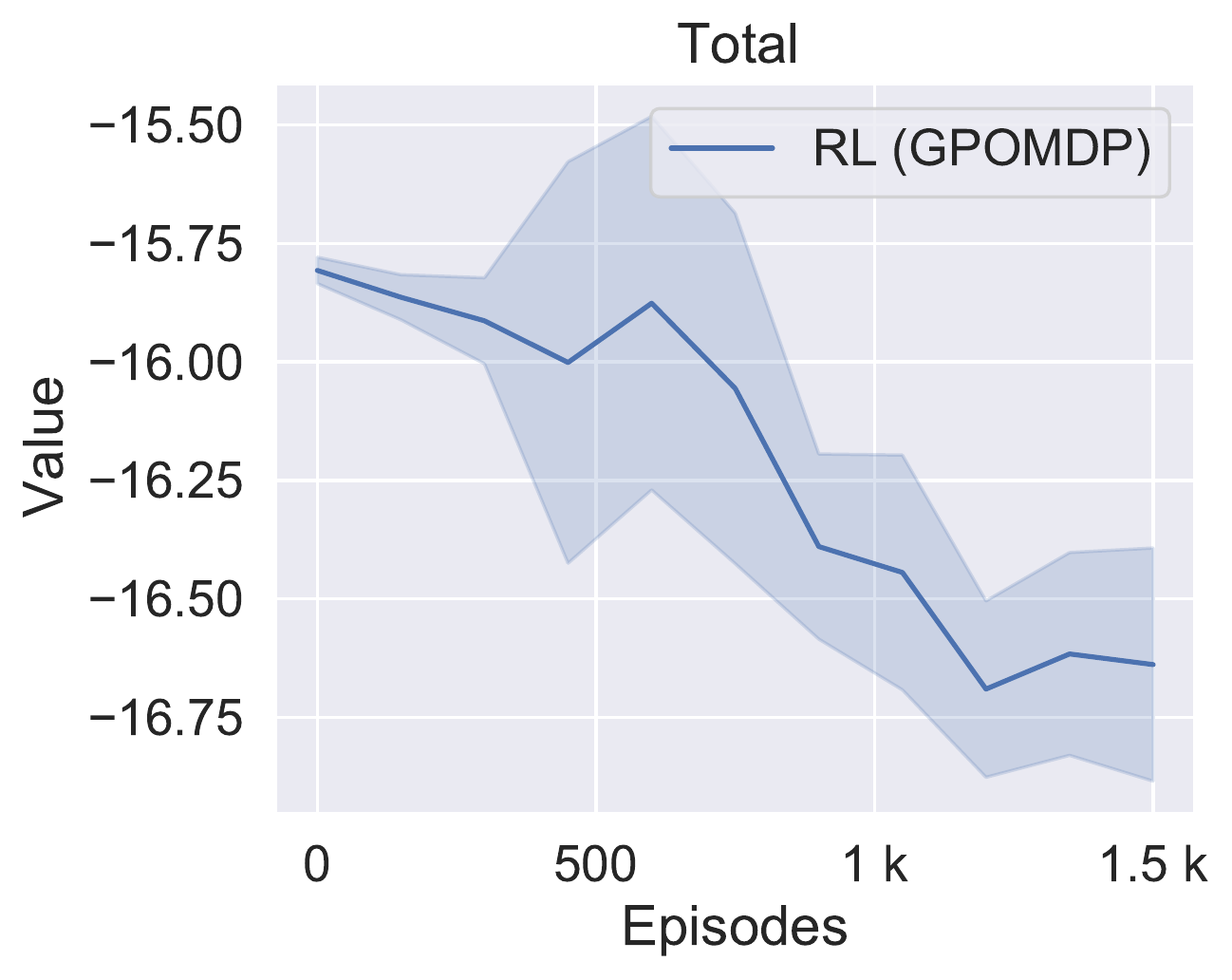}
    }
    }
\end{figure}

Figure \ref{fig:training} shows the evolution of the learning process of the \gls*{rl} recomputation policy evaluated every 150 episodes on the test set. It starts out emulating the \gls*{mpc} strategy achieving similar objective value scores, and from there identifies instances where similar control performance can be achieved without recomputing the \gls*{mpc} solution. It converges after around 1500 episodes corresponding to about 4 hours of real world training time for the cart pendulum system and 400 hours for the battery storage systems. While this is considerable training time, in our experiments the policy generally monotonically improves on the objective and the system could be designed such that even the maximum recomputation interval guarantees safe operation.

\begin{figure}[htbp]
\floatconts
    {fig:eval}
    {\caption{Undiscounted return $G$ over the test sets for the different recomputation policies. The \gls*{rl} bars are mean values with the lines representing one standard deviation.}}
    {%
    \subfigure[Cart pendulum system]{%
        \label{fig:eval:cartpend}
        \includegraphics[width=0.49\textwidth]{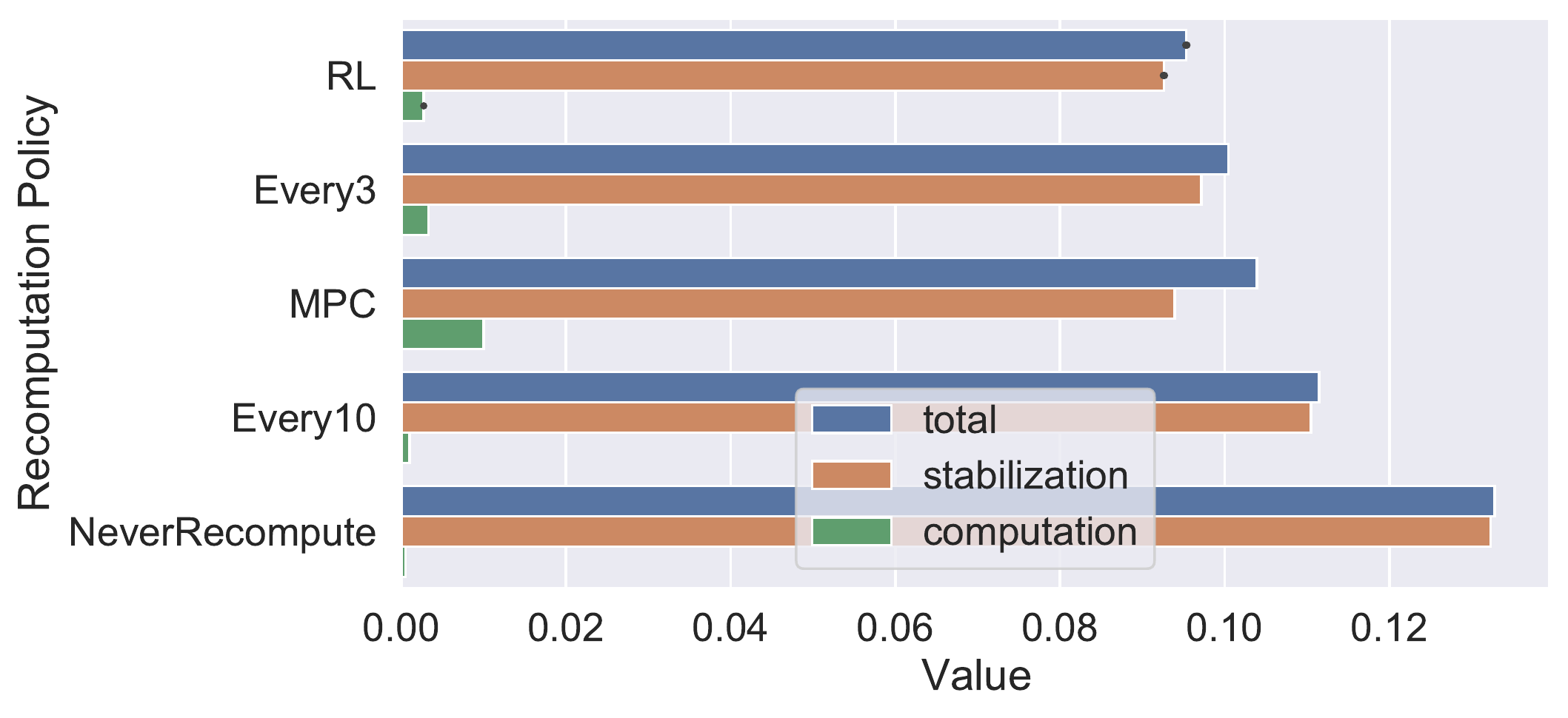}
    }
    \subfigure[Battery storage system]{%
        \label{fig:eval:battery}
        \includegraphics[width=0.49\textwidth]{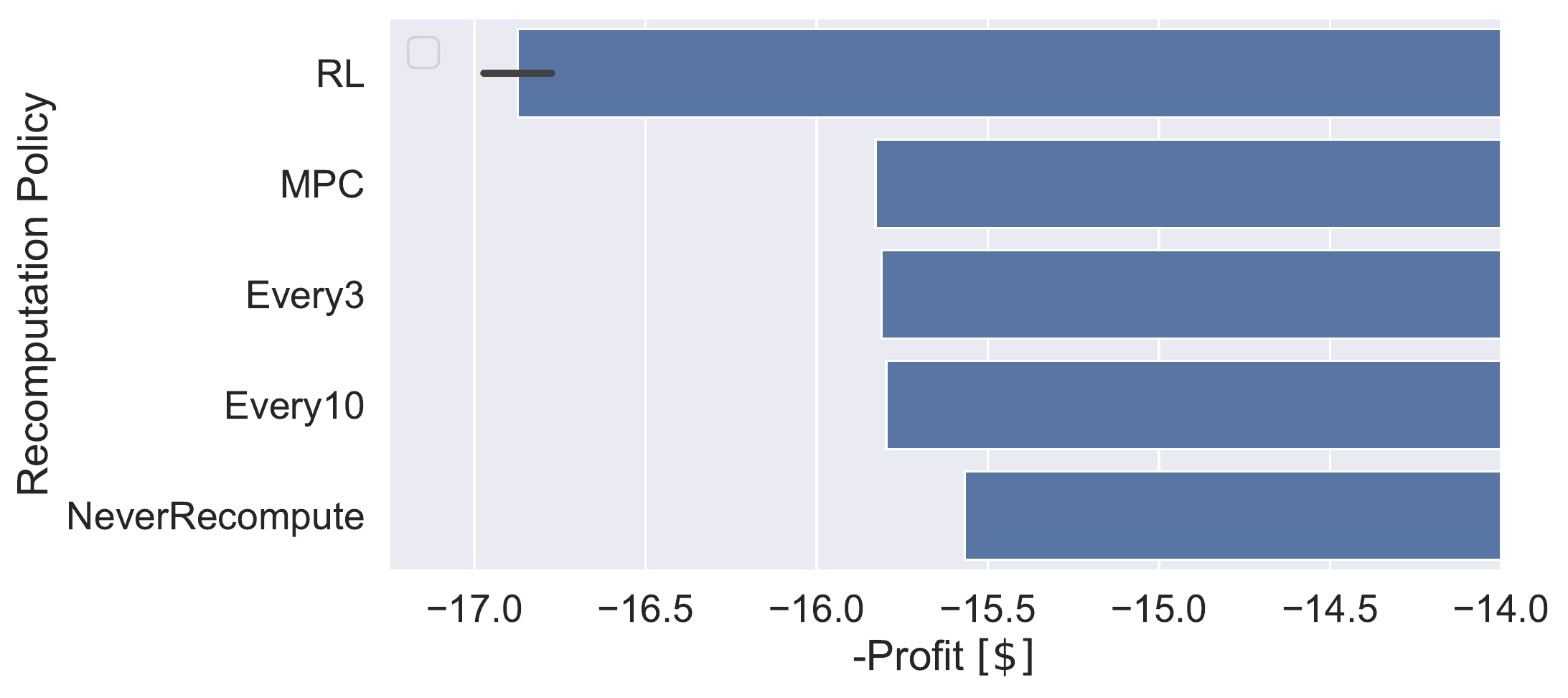}
    }
    }
\end{figure}

The trained \gls*{rl} policy is compared to the baseline controllers in Figure \ref{fig:eval}. It is the best performing policy for both systems, outperforming the second best policy by around 5\% and 7\% for the cart pendulum and battery storage systems, respectively. While this might seem like a minuscule improvement we would argue that the optimization landscape in the recomputation optimization problem is very noisy --- involving complex interactions between the \gls*{mpc} and \gls*{lqr} control laws as well as the plant dynamics --- and the ability to consistently estimate a gradient even with such little room for improvement speaks to the potential of the proposed architecture. For the cart pendulum problem the \gls*{rl} policy is able to have virtually the same control performance as the standard \gls*{mpc} approach with less than one fourth of the computational cost. A noteworthy result is that for the battery system the computational cost is negligible, meaning the performance improvements of the \gls*{rl} policy mainly stem from it recognizing situations where carrying out larger parts of a single \gls*{mpc} solution is more optimal than individual parts of separate solutions. These situations arises due to the battery storage system having considerable stochasticity which is not modeled by the \gls*{mpc}, meaning its solutions are generally suboptimal. 

One major effect we do not account for in this work is the time it takes to obtain the \gls*{mpc} solution, which is often significant compared to the sampling time. The recomputation policy could therefore instead by trained as a self-triggering policy predicting the number of time steps the current solution is viable for. The \gls*{mpc} could then be scheduled so that a new solution is ready when needed, while also giving the \gls*{mpc} more optimization time resulting in a higher quality solution.


\section{Conclusion}\label{sec:conclusion}
In this paper we have presented a novel control system architecture which promises to deliver the performance of  \gls*{mpc} with reduced computational frequency and energy usage, opening up new applications particularly for battery driven systems and other applications where energy is a limiting factor. It is important to note that we did not spend significant time tuning the parameters of the \gls*{mpc} and \gls*{lqr} controllers, being satisfied with establishing the \gls*{mpc} as the superior controller, and employed a basic classical \gls*{rl} algorithm with a simple function approximator containing only a handful of learned parameters. All this is to say that the architecture seems to be fairly robust towards tuning and suboptimality of the individual components, and larger improvements could likely be obtained with more effort put into hyperparameter optimization. Moreover it is not clear how one would even tune the \gls*{lqr} when its objective is to extend the viability of the \gls*{mpc} solution, aside from trial-and-error. One could therefore extend the role of \gls*{rl} in this architecture to also optimize and tune the other controllers, or even have it act as the simpler controller itself. This is left for future work.
\acks{This work was  financed  by grants from the Research Council of Norway (PhD  Scholarships  at  SINTEF grant no. 272402, and NTNU AMOS grant no. 223254).}

\bibliography{references}

\begin{thebibliography}{27}
\providecommand{\natexlab}[1]{#1}
\providecommand{\url}[1]{\texttt{#1}}
\expandafter\ifx\csname urlstyle\endcsname\relax
  \providecommand{\doi}[1]{doi: #1}\else
  \providecommand{\doi}{doi: \begingroup \urlstyle{rm}\Url}\fi

\bibitem[Andersson et~al.(2019)Andersson, Gillis, Horn, Rawlings, and
  Diehl]{Andersson2019}
Joel A~E Andersson, Joris Gillis, Greg Horn, James~B Rawlings, and Moritz
  Diehl.
\newblock {CasADi} -- {A} software framework for nonlinear optimization and
  optimal control.
\newblock \emph{Mathematical Programming Computation}, 11\penalty0
  (1):\penalty0 1--36, 2019.
\newblock \doi{10.1007/s12532-018-0139-4}.

\bibitem[Baumann et~al.(2018)Baumann, Zhu, Martius, and
  Trimpe]{baumann_deep_2018}
Dominik Baumann, Jia-Jie Zhu, Georg Martius, and Sebastian Trimpe.
\newblock Deep reinforcement learning for event-triggered control.
\newblock In \emph{Proceedings of the 57th IEEE International Conference on
  Decision and Control (CDC)}, pages 943--950, December 2018.

\bibitem[Baxter and Bartlett(2001)]{Baxter_2001}
J.~Baxter and P.~L. Bartlett.
\newblock Infinite-horizon policy-gradient estimation.
\newblock \emph{Journal of Artificial Intelligence Research}, 15:\penalty0
  319–350, Nov 2001.
\newblock ISSN 1076-9757.
\newblock \doi{10.1613/jair.806}.

\bibitem[Bemporad et~al.(2002)Bemporad, Morari, Dua, and
  Pistikopoulos]{bemporad_explicit_2002}
Alberto Bemporad, Manfred Morari, Vivek Dua, and Efstratios~N. Pistikopoulos.
\newblock The explicit linear quadratic regulator for constrained systems.
\newblock \emph{Automatica}, 38\penalty0 (1):\penalty0 3 -- 20, 2002.
\newblock ISSN 0005-1098.
\newblock \doi{https://doi.org/10.1016/S0005-1098(01)00174-1}.

\bibitem[Berglind et~al.(2012)Berglind, Gommans, and
  Heemels]{berglind_self_2012}
J.D.J.~Barradas Berglind, T.M.P. Gommans, and W.P.M.H. Heemels.
\newblock Self-triggered mpc for constrained linear systems and quadratic
  costs.
\newblock \emph{IFAC Proceedings Volumes}, 45\penalty0 (17):\penalty0 342 --
  348, 2012.
\newblock ISSN 1474-6670.
\newblock \doi{https://doi.org/10.3182/20120823-5-NL-3013.00058}.
\newblock 4th IFAC Conference on Nonlinear Model Predictive Control.

\bibitem[Berner et~al.(2019)Berner, Brockman, Chan, Cheung, Debiak, Dennison,
  Farhi, Fischer, Hashme, Hesse, et~al.]{berner2019dota}
Christopher Berner, Greg Brockman, Brooke Chan, Vicki Cheung, Przemys{\l}aw
  Debiak, Christy Dennison, David Farhi, Quirin Fischer, Shariq Hashme, Chris
  Hesse, et~al.
\newblock Dota 2 with large scale deep reinforcement learning.
\newblock \emph{arXiv preprint arXiv:1912.06680}, 2019.

\bibitem[{Cagienard} et~al.(2004){Cagienard}, {Grieder}, {Kerrigan}, and
  {Morari}]{cagienard_move_2004}
R.~{Cagienard}, P.~{Grieder}, E.~C. {Kerrigan}, and M.~{Morari}.
\newblock Move blocking strategies in receding horizon control.
\newblock In \emph{2004 43rd IEEE Conference on Decision and Control (CDC)
  (IEEE Cat. No.04CH37601)}, volume~2, pages 2023--2028 Vol.2, 2004.
\newblock \doi{10.1109/CDC.2004.1430345}.

\bibitem[{Chakrabarty} et~al.(2018){Chakrabarty}, {Zavitsanou}, {Doyle}, and
  {Dassau}]{chakrabarty_event_2018}
A.~{Chakrabarty}, S.~{Zavitsanou}, F.~J. {Doyle}, and E.~{Dassau}.
\newblock Event-triggered model predictive control for embedded artificial
  pancreas systems.
\newblock \emph{IEEE Transactions on Biomedical Engineering}, 65\penalty0
  (3):\penalty0 575--586, 2018.
\newblock \doi{10.1109/TBME.2017.2707344}.

\bibitem[Feng et~al.(2012)Feng, Gutvik, Johansen, Sui, and
  Brubakk]{le_approximate_2012}
Le~Feng, Christian Gutvik, Tor Johansen, Dan Sui, and Alf Brubakk.
\newblock Approximate explicit nonlinear receding horizon control for
  decompression of divers.
\newblock \emph{Control Systems Technology, IEEE Transactions on}, 20:\penalty0
  1275--1284, 09 2012.
\newblock \doi{10.1109/TCST.2011.2162516}.

\bibitem[Goebel and Allgöwer(2015)]{goebel_simple_2015}
Gregor Goebel and Frank Allgöwer.
\newblock A simple semi-explicit mpc algorithm.
\newblock \emph{IFAC-PapersOnLine}, 48\penalty0 (23):\penalty0 489 -- 494,
  2015.
\newblock ISSN 2405-8963.
\newblock \doi{https://doi.org/10.1016/j.ifacol.2015.11.326}.
\newblock 5th IFAC Conference on Nonlinear Model Predictive Control NMPC 2015.

\bibitem[Gondhalekar et~al.(2015)Gondhalekar, Dassau, and {Doyle
  III}]{gondhalekar_tackling_2015}
Ravi Gondhalekar, Eyal Dassau, and Francis~J. {Doyle III}.
\newblock Tackling problem nonlinearities \& delays via asymmetric,
  state-dependent objective costs in mpc of an artificial pancreas.
\newblock \emph{IFAC-PapersOnLine}, 48\penalty0 (23):\penalty0 154 -- 159,
  2015.
\newblock ISSN 2405-8963.
\newblock \doi{https://doi.org/10.1016/j.ifacol.2015.11.276}.
\newblock 5th IFAC Conference on Nonlinear Model Predictive Control NMPC 2015.

\bibitem[Gros et~al.(2020)Gros, Zanon, and Bemporad]{gros_safe_2020}
Sebastien Gros, Mario Zanon, and Alberto Bemporad.
\newblock Safe reinforcement learning via projection on a safe set: how to
  achieve optimality?
\newblock In \emph{IFAC 2020}, 02 2020.

\bibitem[Heemels et~al.(2008)Heemels, Sandee, and Bosch]{heemels_analysis_2008}
W.~P. M.~H. Heemels, J.~H. Sandee, and P.~P. J. Van~Den Bosch.
\newblock Analysis of event-driven controllers for linear systems.
\newblock \emph{International Journal of Control}, 81\penalty0 (4):\penalty0
  571--590, 2008.
\newblock \doi{10.1080/00207170701506919}.

\bibitem[{Iino} et~al.(2009){Iino}, {Hatanaka}, and {Fujita}]{iino_event_2009}
Y.~{Iino}, T.~{Hatanaka}, and M.~{Fujita}.
\newblock Event-predictive control for energy saving of wireless networked
  control system.
\newblock In \emph{2009 American Control Conference}, pages 2236--2242, 2009.
\newblock \doi{10.1109/ACC.2009.5160732}.

\bibitem[Li and Shi(2014)]{li_event_2014}
Huiping Li and Yang Shi.
\newblock Event-triggered robust model predictive control of continuous-time
  nonlinear systems.
\newblock \emph{Automatica}, 50\penalty0 (5):\penalty0 1507 -- 1513, 2014.
\newblock ISSN 0005-1098.
\newblock \doi{https://doi.org/10.1016/j.automatica.2014.03.015}.

\bibitem[Lucia et~al.(2017)Lucia, T{\u{a}}tulea-Codrean, Schoppmeyer, and
  Engell]{lucia2017rapid}
Sergio Lucia, Alexandru T{\u{a}}tulea-Codrean, Christian Schoppmeyer, and
  Sebastian Engell.
\newblock Rapid development of modular and sustainable nonlinear model
  predictive control solutions.
\newblock \emph{Control Engineering Practice}, 60:\penalty0 51--62, 2017.

\bibitem[Mnih et~al.(2015)Mnih, Kavukcuoglu, Silver, Rusu, Veness, Bellemare,
  Graves, Riedmiller, Fidjeland, Ostrovski, et~al.]{mnih2015human}
Volodymyr Mnih, Koray Kavukcuoglu, David Silver, Andrei~A Rusu, Joel Veness,
  Marc~G Bellemare, Alex Graves, Martin Riedmiller, Andreas~K Fidjeland, Georg
  Ostrovski, et~al.
\newblock Human-level control through deep reinforcement learning.
\newblock \emph{nature}, 518\penalty0 (7540):\penalty0 529--533, 2015.

\bibitem[Peters and Schaal(2008)]{peters_reinforcement_2008}
Jan Peters and Stefan Schaal.
\newblock Reinforcement learning of motor skills with policy gradients.
\newblock \emph{Neural Networks}, 21\penalty0 (4):\penalty0 682 -- 697, 2008.
\newblock ISSN 0893-6080.
\newblock \doi{https://doi.org/10.1016/j.neunet.2008.02.003}.
\newblock Robotics and Neuroscience.

\bibitem[Schrittwieser et~al.(2019)Schrittwieser, Antonoglou, Hubert, Simonyan,
  Sifre, Schmitt, Guez, Lockhart, Hassabis, Graepel, Lillicrap, and
  Silver]{Schrittwieser2019MasteringAG}
Julian Schrittwieser, Ioannis Antonoglou, T.~Hubert, K.~Simonyan, L.~Sifre,
  S.~Schmitt, A.~Guez, Edward Lockhart, Demis Hassabis, T.~Graepel,
  T.~Lillicrap, and D.~Silver.
\newblock Mastering atari, go, chess and shogi by planning with a learned
  model.
\newblock \emph{ArXiv}, abs/1911.08265, 2019.

\bibitem[{Scokaert} et~al.(1999){Scokaert}, {Mayne}, and
  {Rawlings}]{scokaert_suboptimal_1999}
P.~O.~M. {Scokaert}, D.~Q. {Mayne}, and J.~B. {Rawlings}.
\newblock Suboptimal model predictive control (feasibility implies stability).
\newblock \emph{IEEE Transactions on Automatic Control}, 44\penalty0
  (3):\penalty0 648--654, 1999.
\newblock \doi{10.1109/9.751369}.

\bibitem[Sedghi et~al.(2020)Sedghi, Ijaz, Noor-A-Rahim, Witheephanich, and
  Pesch]{sedghi2020machine}
Leila Sedghi, Zohaib Ijaz, Md. Noor-A-Rahim, Kritchai Witheephanich, and Dirk
  Pesch.
\newblock Machine learning in event-triggered control: Recent advances and open
  issues, 2020.

\bibitem[Sutton and Barto(2018)]{sutton_reinforcement_2018}
Richard~S. Sutton and Andrew~G. Barto.
\newblock \emph{Reinforcement Learning: An Introduction}.
\newblock A Bradford Book, Cambridge, MA, USA, 2018.
\newblock ISBN 0262039249.
\newblock \doi{10.5555/3312046}.

\bibitem[Wächter and Biegler(2006)]{ipopt}
Andreas Wächter and Lorenz Biegler.
\newblock On the implementation of an interior-point filter line-search
  algorithm for large-scale nonlinear programming.
\newblock \emph{Mathematical programming}, 106:\penalty0 25--57, 03 2006.
\newblock \doi{10.1007/s10107-004-0559-y}.

\bibitem[{Yang} et~al.(2019){Yang}, {He}, and {Liu}]{yang_event_2019}
X.~{Yang}, H.~{He}, and D.~{Liu}.
\newblock Event-triggered optimal neuro-controller design with reinforcement
  learning for unknown nonlinear systems.
\newblock \emph{IEEE Transactions on Systems, Man, and Cybernetics: Systems},
  49\penalty0 (9):\penalty0 1866--1878, 2019.
\newblock \doi{10.1109/TSMC.2017.2774602}.

\bibitem[{Yoo} and {Johansson}(2019)]{yoo_event_2019}
J.~{Yoo} and K.~H. {Johansson}.
\newblock Event-triggered model predictive control with a statistical learning.
\newblock \emph{IEEE Transactions on Systems, Man, and Cybernetics: Systems},
  pages 1--11, 2019.
\newblock \doi{10.1109/TSMC.2019.2916626}.

\bibitem[Zanon and Gros(2020)]{zanon_safe_2020}
Mario Zanon and Sebastien Gros.
\newblock Safe reinforcement learning using robust mpc.
\newblock \emph{IEEE Transactions on Automatic Control}, PP:\penalty0 1--1, 09
  2020.
\newblock \doi{10.1109/TAC.2020.3024161}.

\bibitem[{Åström} and {Bernhardsson}(2002)]{astrom_comparison_2002}
K.~J. {Åström} and B.~M. {Bernhardsson}.
\newblock Comparison of riemann and lebesgue sampling for first order
  stochastic systems.
\newblock In \emph{Proceedings of the 41st IEEE Conference on Decision and
  Control, 2002.}, volume~2, pages 2011--2016 vol.2, 2002.
\newblock \doi{10.1109/CDC.2002.1184824}.

\end{thebibliography}

\end{document}